\documentstyle[prb,preprint,aps]{revtex}
\newcommand{\be}{\begin{eqnarray}}
\newcommand{\ee}{\end{eqnarray}}
\newcommand{\OP}{\psi}

\newcommand{\F}{{\cal F}}

\begin{document}

\title{Perturbation Expansion in Phase Ordering Kinetics}
\author{Gene F. Mazenko }
\address{The James Franck Institute and the Department of Physics\\
The University of Chicago\\
Chicago, Illinois 60637}
\date{\today}
\maketitle
%

%
%
\begin{abstract}

A consistent perturbation theory expansion is presented for
phase ordering kinetics in the case of a nonconserved scalar order
parameter.  At lowest order in this formal expansion one obtains the
theory due to Ohta,  Jasnow and Kawasaki (OJK).  At next order,
worked out explicitly in d dimensions, one has small corrections
to the OJK result for
the nonequilibrium exponent $\lambda$ and the introduction of a new
exponent $\nu$ governing the algebraic component of the decay of the
order parameter scaling function at large scaled distances.

\end{abstract}


Significant progress has been made on the theory of phase ordering
kinetics\cite{2} using methods that introduce auxiliary fields that are
taken to have gaussian
statistics.  The methods developed in
OJK\cite{OJK} and TUG\cite{TUG} each have appealing aspects and
separately give
good descriptions of different aspects
of the ordering problem.
A major lingering
question is why these methods work as well as they do and how they can
be reconciled and improved.  Thus there has been a search\cite{BH,post}
for the field
theory description where OJK or TUG is the zeroth order approximation
in some {\bf systematic} expansion.  Such an expansion is presented in
this paper.  It has the OJK result as its zeroth order approximation.

The system studied here is the domain wall dynamics generated by the
time-dependent Ginzburg-Landau
model satisfied by a nonconserved scalar order
parameter $\psi (\vec{r},t)$:
\be
\Lambda (1) \psi (1)=-V'[\psi (1)]
\ee
where, in dimensionless units,
$\Lambda (1)=\frac{\partial}{\partial t_{1}}-\nabla^{2}_{1}$
is the diffusion operator and
$1$ denotes
$({\bf r_{1}},t_{1})$.
We expect our main results to hold
for any
symmetric
degenerate double-well potential $V[\psi ]$ and to be
independent of the exact nature of
the initial and final states, provided they  are disordered and ordered
states respectively.  Therefore it is convenient to quench to zero
temperature and set any noise term to zero.

It is well
established\cite{2} that for late times following a quench
from the disordered to the
ordered phase the
order-parameter
correlation function has a  scaling form
$C(12) \equiv \langle \OP (1) \OP (2) \rangle
= \psi_{0}^{2} \F(x,t_{1}/t_{2})$
where $\psi_{0}$ is the magnitude of the order-parameter in the
ordered phase and
the scaled length $x$ is defined as
$\vec{x} = (\vec{r}_{1}-\vec{r}_{2})/L$ where
$L=\sqrt{4T}\equiv\sqrt{2(t_{1}+t_{2})}$ is a growing
characteristic length. In the perturbation theory
developed here we find that
\be
\F(x,t_{1}/t_{2})=\frac{2}{\pi}~sin^{-1}
\left[~\Phi(t_{1}/t_{2}) f(x,t_{1}/t_{2})\right]
\label{eq:3}
\ee
with the normalizations $f(0,t_{1}/t_{2})=1$ and $\Phi (1)=1$.  The  quantity
$\Phi(t_{1}/t_{2})$, related to the
on-site order parameter autocorrelation function, has the form\cite{quan}
\be
\Phi (t_{1}/t_{2})=\left(\frac{\sqrt{t_{1}t_{2}}}{T}\right)^{\lambda}
\left( 1 +\Delta\Phi (t_{1}/t_{2})\right)
\label{eq:4}
\ee
where $\Delta\Phi (t_{1}/t_{2})$ is a regular function of $t_{1}/t_{2}$
which is of second order\cite{second} in the expansion.
The nonequilibrium\cite{FH} exponent $\lambda$, which gives the decay for
$t_{1}\gg t_{2}$ and vice versa,  is given to second order by
\be
\lambda =\frac{d}{2}+\omega^{2}\frac{2^{d}M_{d}}{3^{d/2+1}} ~~~.
\ee
The dimensionality dependent quantities, $\omega $, $K_{d}$ and
$M_{d}$,  are given by
\be
2\omega+\omega^{2}2^{d}\left(K_{d}+
\frac{M_{d}}{3^{d/2+1}}\right)=1+\frac{d}{2}
\label{eq:6}
\ee
\be
K_{d}=\int_{0}^{1}dz \frac{z^{d/2-1}}{[(1+z)(3-z)]^{d/2}}
\label{eq:7}
\ee
and
\be
M_{d}=\int_{0}^{1}dz \frac{z^{d/2-1}}{[1+z]^{d}}=
\frac{1}{2}\frac{\Gamma^{2}(d/2)}{\Gamma (d)} ~~~.
\label{eq:8}
\ee
$f(x,t_{1}/t_{2})$ can be written in the form
\be
f(x,t_{1}/t_{2})=\frac{f_{0}(x)}{(1+x^{2})^{\nu /2}}
\left[1+\Delta f(x,t_{1}/t_{2})\right]
\label{eq:9}
\ee
where $\Delta f(x,t_{1}/t_{2})$ is a regular function of its arguments and
is second order in the expansion and $f_{0}(x)=e^{-\frac{1}{2}x^{2}}$
is just the OJK result for the scaling function. The
exponent $\nu$ governing the algebraic component\cite{TUG1} of the large
$x$ behavior, shared by $f$ and $\F$, is given by
\be
\nu =\omega^{2}2^{d+1}\left(
K_{d}+\frac{M_{d}}{3^{d/2+1}}\right) ~~~.
\ee
At lowest order one has the
OJK results $\lambda =\frac{d}{2}$ and $\nu =0$.  While $K_{d}$ and
$\omega$ can be worked out analytically for specific values of $d$,
the expressions are not very illuminating.  Numerically we have for
$d=1$, $\omega =0.4601..$, $\lambda =0.6268..$, $\nu =1.1596..$;
$d=2$, $\omega =0.6877..$, $\lambda =1.1051..$, $\nu =1.2492..$;
$d=3$, $\omega =0.9067..$, $\lambda =1.5828..$, $\nu =1.3732..$.
For large $d$, $\omega\rightarrow \frac{d}{2}(\sqrt{2}-1)$,
$\lambda\rightarrow \frac{d}{2}$, and
$\nu\rightarrow d(3-2\sqrt{2})$.  The {\it best} estimates\cite{LM}
are $d=1$, $\lambda =1$,
$d=2$, $\lambda =1.246\pm 0.02$, and
$d=3$, $\lambda =1.838\pm 0.2$.
For equal times $t_{1}=t_{2}$,
$\F(x,1)=\F(x)=\frac{2}{\pi}~sin^{-1}f(x,1)$.
For small $x$, where $f$ is analytic in an expansion in even powers of $x$,
$\F (x)$ satisfies Porod's law\cite{porod}
and the Tomita\cite{tomita} sum rule.

The perturbative treatment developed here is
a two step process.  The first step is to introduce
an auxiliary field $m(1)$ which
satisfies an equation of motion
\be
\Lambda (1)\nabla_{i}m(1)=2\xi (t_{1}) \rho_{i}(1)
\equiv 2\xi (t_{1})\delta (m(1))\nabla_{i}m(1)~~~.
\label{eq:12}
\ee
The quantity $\xi (t_{1})$ must be chosen to go as $L^{-1}(t_{1})$ for large
times with a coefficient which must be adjusted if the system is to order,
much like the temperature must be set to $T_{c}$ in critical phenomena.
This equation of motion has the appealing feature that $ \nabla_{i}m(1)$ is
a diffusion field driven by a
source $\rho_{i}(1) $ which
is the invariant density\cite{conservation} of domain walls for the system.
We show below that we can organize a perturbation theory expansion for this
field theory about a gaussian zeroth order theory.

The second step in the analysis is to show that one can map the phase
ordering kinetics problem onto this new field theory.  This mapping is
via the transformation
$\psi =\sigma [m] + u[m]$
where $\sigma [m]$ satisfies the Euler-Lagrange equation\cite{TUG2}
for the associated
stationary interface problem with $m$ the coordinate:
$\sigma_{2}=V'[\sigma [m]]$
where we introduce
the notation $\sigma_{\ell}=\frac{d^{\ell}\sigma }{dm^{\ell}}$.
The idea is that $m$ grows large away from an interface and
$\sigma \rightarrow \psi_{0}sgn (m)$ in the {\bf bulk}.
Inserting this mapping into the equation of motion
for $\psi$ we find:
\be
\Lambda (1)u(1)+\sigma_{1}(1)\Lambda (1)m(1)=-V'[\sigma (1)+u(1) ]
+\sigma_{2}(1)(\nabla m(1))^{2} ~~~.
\ee
The most important property of the solution of this equation
for $u[m]$ is that
$lim_{|m|\rightarrow \infty}u [m]=0$
and $\psi\rightarrow \psi_{0} sgn~m$.
In the regime where $|m|$ is
large, $\sigma^{2}\approx \psi_{0}^{2}$, the derivatives of $\sigma $ go
exponentially to zero, the linearized equation for
$u$ shows that it has
acquired a {\it mass}
and, in the long-time long-distance limit,
u is seen to go exponentially to zero.
The universal properties of interest are associated with
bulk quantities like
$C(12...n)=<\psi (1)\psi (2)\cdots \psi (n)>$
where the points $\vec{r}_{1},\vec{r}_{2},...,\vec{r}_{n}$
are not constrained to be close together.
Because the field $u(1)$ is nonzero only near
interfaces the average
$<u(1)\sigma (2)\cdots >$ is down by a factor of
$1/L^{2}(t_{1})$ relative to
$<\sigma (1)\sigma (2)\cdots >$.  Thus if we focus on these properties
we do not need to know $u$ in detail.

We will develop the field theory generated by
the equation of motion, Eq.(\ref{eq:12}),
for $m(1)$
using the standard
Martin-Siggia-Rose\cite{MSR} method in its functional integral form as formulated by
DeDominicis and Peliti\cite{DP}.
In the MSR method the field
theoretical development requires a doubling of fields to include a
field $\hat{m}_{i}$ conjugate to $\nabla_{i}m$.
We also organize things so that the initial field $m_{0}(\vec{r})$ is
also treated as independent.  Following
standard procedures, averages of interest are given as functional
integrals over the
fields $m$, $\hat{m}_{i}$ and $m_{0}$ weighted by an action
$A$:
\be
A(m,\hat{m},m_{0})=i\int d1\sum_{i=1}^{d}\hat{m}_{i}(1)\left[\Lambda (1)\nabla_{i}
m(1)-2\xi(1)\rho_{i}(1)\right]
\ee
\be
-i\int d1\sum_{i=1}^{d}\hat{m}_{i}(1)\delta (t_{1}-t_{0})
\nabla_{i}m_{0}(1)
-\frac{1}{2}\int d^{d}r_{1}\int d^{d}r_{2}~m_{0}(\vec{r}_{1})
g^{-1}(\vec{r}_{1}-\vec{r}_{2})m_{0}(\vec{r}_{2})
\nonumber
\ee
where we use the notation
$d1=dt_{1}d^{d}r_{1}$
and where are assume, as is appropriate in this case, that the initial field
is gaussian and has a variance given by
$<m_{0}(\vec{r}_{1})m_{0}(\vec{r}_{2})>=g(\vec{r}_{1}-\vec{r}_{2})$ .
We will not have to be very specific about the form of the initial
correlation function $g$.  It is convenient to generate our
correlation functions as functional derivatives in terms of sources,
$h(1)$ and $\hat{h}(1)$, which
couple to the conjugate fields.  Thus we introduce the total action
$A_{T}$:
\be
A_{T}=A+\int d1 \left[h(1)m(1)
+\hat{h}(1)\sum_{i=1}^{d}\nabla_{i}\hat{m}_{i}(1)\right] ~~~.
\ee
The fundamental equations of motion are given then by
\be
<\frac{\delta}{\delta \Psi (1)}A_{T}(m,\hat{m},m_{0})>_{h}=0
\label{eq:16}
\ee
where $\Psi (1)=\{m(1),\hat{m}_{i}(1),m_{0}(1)\}$ and
the averages are over the {\it total} action $A_{T}$.  Taking the
functional derivatives of $A_{T}$ and and, using Eq.(\ref{eq:16}) with
$\Psi (1)=m_{0}(1)$ to eliminate the initial field
in terms of the hatted field, we obtain the two basic equations
\be
-i\left[\tilde{\Lambda}(1)<\hat{M}(1)>_{h}
+2\xi(1)<\hat{\rho}(1)>_{h}
\right]=h(1)
\label{eq:17}
\ee
\be
i\left[\Lambda (1)\nabla_{i}
<m(1)>_{h}-2\xi(1)<\rho_{i}(1)>_{h}\right]
=-\nabla_{i}^{(1)}\int d2 ~\Pi_{0} (12)<\hat{M}(2)>_{h}+\nabla_{i}\hat{h}(1)
{}.
\label{eq:18}
\ee
Here we have introduced the quantities
$\Pi_{0} (12) \equiv \delta (t_{1}-t_{0})\delta (t_{1}-t_{2})
g(\vec{r}_{1}-\vec{r}_{2}) $
,$\tilde{\Lambda}(1)=\frac{\partial}{\partial t_{1}}+\nabla^{2}_{1}$,
$\hat{\rho}(1)=\delta (m(1))\hat{M}(1)$,
and
$\hat{M}(1)=\sum_{i=1}^{d}\nabla_{i}\hat{m}_{i}(1) $.
Notice that the fundamental equations Eqs.(\ref{eq:17})  and
(\ref{eq:18}) depend on $\hat{M}$
and not $\hat{m}_{i}$ and we have constructed $\hat{h}$ such that it
couples to $\hat{M}$.  All correlation functions of interest can be
generated as functional derivatives of $<m(1)>_{h}$ or $<\hat{M}(1)>_{h}$
with respect to $h$ and $\hat{h}$.

Taking the functional derivative of
Eqs.(\ref{eq:17})  and  (\ref{eq:18}) with respect to $h(2)$
gives the equations for the two point response and correlation
functions:
\be
-i\left[\tilde{\Lambda}(1)G_{\hat{M}m}(12)
+\hat{Q}_{2}(12)
\right]=\delta (12)
\label{eq:19}
\ee
\be
i\left[\Lambda (1)\nabla_{i}^{(1)}G_{mm}(12)
-Q_{2}^{(i)}(12)
\right]
=-\nabla_{i}^{(1)}\int d3 ~\Pi_{0} (13)G_{\hat{M}m}(32) ~~~.
\label{eq:20}
\ee
where we have used
\be
\hat{Q}_{n}(12...n)=
\frac{\delta^{n-1}}{\delta h(n)\delta h(n-1)\cdots\delta h(2)}
\left[2\xi(1)<\hat{\rho}(1)>_{h}
\right]
\nonumber
\ee
\be
Q_{n}^{(i)}(12...n)=
\frac{\delta^{n-1}}{\delta h(n)\delta h(n-1)\cdots\delta h(2)}
\left[2\xi(1)<\rho_{i}(1)>_{h}\right]~~~,
\nonumber
\ee
evaluated for $n=2$ and introduced the notation
\be
G_{\hat{M}mm..m}(123..n)=\frac{\delta^{n-1}<\hat{M}(1)>_{h}}{\delta\hat{h}(2)
\delta h(3)\cdots\delta h(n)}
\nonumber
\ee
\be
G_{n}(12\cdots n)=\frac{\delta^{n-1}}{\delta h(n)\delta h(n-1)
\cdots \delta h(2)}<m(1)>_{h} ~~~.
\nonumber
\ee
The functional derivative of Eq.(\ref{eq:18}) with respect to
$\hat{h}(2)$ is redundant
compared with Eq.(\ref{eq:19})
because of the relation
$G_{m\hat{M}}(12)=G_{\hat{M}m}(21)$ .
The equations governing the $n^{th}$ order cumulants are given by
\be
-i\left[\tilde{\Lambda}(1)G_{\hat{M}m...m}(12...n)
+\hat{Q}_{n}(12...n)\right]=0
\nonumber
\ee
\be
i\left[\Lambda (1)\nabla_{i}G_{n}(12...n)-Q_{n}^{(i)}(12...n)\right]
=-\nabla_{i}^{(1)}\int
d\bar{1} ~\Pi_{0} (1\bar{1})G_{\hat{M}m...m}(\bar{1}2...n) ~~~.
\nonumber
\ee

The point now is to show that there is a consistent perturbation expansion
for this theory.
After introducing
the one-point
probability distribution
$P_{h}(x_{1},1)=<\delta (x_{1}-m(1))>_{h}$,
it is easy to  show
\be
Q_{1}^{(i)}(1)=2\xi (1)<\rho_{i}(1)>_{h
}=\nabla_{i}^{(1)}\int dx_{1}\xi (1) sgn (x_{1})
P_{h}(x_{1},1)
\label{eq:29}
\ee
\be
\hat{Q}_{1}(1)
=2\xi (1)<\hat{\rho}(1)>_{h
}=2\xi (1)
\left[<\hat{M}(1)>_{h}
+\frac{\delta}{\delta \hat{h}(1)}\right]
P_{h}(0,1)~~~.
\label{eq:30}
\ee
Then any perturbation theory expansion for $P_{h}(x_{1},1)$ will
lead immediately to an expansion for
$Q_{n}^{(i)}$ and $\hat{Q}_{n}$ by functional
differentiation.
The expansion for $P_{h}(x_{1},1)$ is straightforward.  Using the
integral representation for the $\delta$-function we have
\be
P_{h}(x_{1},1)= \int\frac{dk_{1}}{2\pi}e^{-ik_{1}x_{1}}
<e^{H(1)}>_{h}
= \int\frac{dk_{1}}{2\pi}e^{-ik_{1}x_{1}}
exp\left[\sum_{n=1}^{\infty}\frac{1}{n!}G_{H}^{(n)}(1)\right]
\nonumber
\ee
where
$H(1)\equiv ik_{1}m(1)$
and
$G_{H}^{(n)}(1)$ is the $n^{th}$ order cumulant for the field $H(1)$.
Since $H(1)$ is proportional to $m(1)$ these are, up to factors
of $ik_{1}$ to the $n^{th}$ power, just the cumulants for the $m$
field.
It turns out that
in zero external field we can take
$G_{n}$ to be of ${\cal O}( \frac{n}{2}-1)$.
Keeping
terms up to the 4-point cumulant it is easy to see that
\be
P_{h}(x_{1},1)=\left[1-\frac{1}{3!}G_{3}(111)\frac{d^{3}}{dx_{1}^{3}}
+\frac{1}{4!}G_{4}(1111)\frac{d^{4}}{dx_{1}^{4}}+\cdots\right]
P_{h}^{(0)}(x_{1},1)
\label{eq:32}
\ee
where
\be
P_{h}^{(0)}(x_{1},1)=\int\frac{dk_{1}}{2\pi} \Phi_{0}(k_{1},1)e^{-ik_{1}x_{1}}
\label{eq:33}
\ee
and
\be
\Phi_{0}(k_{1},1)=e^{-ik_{1}(x_{1}-G_{1}(1))}
e^{-\frac{1}{2}k_{1}^{2}G_{2}(11)} ~~~.
\label{eq:34}
\ee
Eqs.(\ref{eq:32}), (\ref{eq:33})  and (\ref{eq:34}) inserted back into
Eqs.(\ref{eq:29}) and (\ref{eq:30})
gives a relation between the $m$-field cumulants and
the quantities $Q_{n}^{(i)}$ and $\hat{Q}_{n}$.  These quantities are
ultimately related to derivatives of $P_{h}^{(0)}[0,1]$.  Because of this
we see that each term in the perturbation theory expansion
in zero external fields
is weighted by factors of
\be
\phi_{n}(1)=\int\frac{dk_{1}}{2\pi}k_{1}^{2n}e^{-\frac{1}{2}k_{1}^{2}S(1)}
=\left(-2\frac{\partial}{\partial S(1)}\right)^{n}\phi_{0}(1)
\ee
where
$\phi_{0}(1)=\frac{1}{\sqrt{2\pi S(1)}}$
and
$S(1)=<m^{2}(1)> $.
The structure of the perturbation theory becomes clear if one looks at
two classes of contributions to
$Q_{2n}^{(i)}$. The first class corresponds to the $2n-1$
derivatives which all act on $\Phi_{0}(k_{1},1)$ to give contributions
to $Q_{2n}^{(i)}(12\cdots ,2n)$ of the form
\be
\nabla_{i}^{(1)}\xi (1)\phi_{n-1}(1)
G_{2}(12)G_{2}(13)\cdots G_{2}(1,2n) ~~~.
\nonumber
\ee
There is another set of contributions where all of the derivatives
except the first acts on the factor multiplying $\Phi_{0}(k_{1},1)$ and gives
a contribution
to $Q_{2n}^{(i)}(12\cdots ,2n)$ of the form
\be
\nabla_{i}^{(1)}\xi (1)\phi_{0}(1)
G_{2n}(12\cdots ,2n)~~~.
\nonumber
\ee
Inserting these results into the expression for the $2n$ point cumulant we
see term by term that the leading order behavior goes as
$G_{2n}\approx \phi_{n-1}$.
Orders in the expansion correspond to  the
sum of the labels n on $\phi_{n}$ in products of $\phi$'s.  Thus a term
with factors
$\phi_{1}\phi_{2}\phi_{1}$, typically associated with different times,
is of ${\cal O}(4)$.
It should be emphasized that at this stage that this is a {\it formal}
expansion.  At order $n$ it is true that
$\phi_{n}\approx L^{-(2n+1)}$ is small, however it will be multiplied ,
depending on the quantity
expanded, by positive factors of $L(t)$ such that each term in the
expansion in $\phi_{n}$ has the same
overall leading power with respect to $L(t)$.

It is then straightforward\cite{causality}
to work out the expansion for the two-point cumulant and obtain at
zeroth order:
\be
\hat{Q}_{2}(12)=\omega_{0} (1)G_{\hat{M}m}(12)
\label{eq:38}
\ee
\be
Q_{2}^{(i)}(12)=\nabla_{i}^{(1)}\omega_{0} (1)G_{2}(12)
\label{eq:39}
\ee
where we have defined
$\omega_{n} (1) =2\xi(1)\phi_{n}(1)$.
The equations of motion at zeroth order for the two point response and
correlation
functions is given by Eqs.(\ref{eq:19}) and (\ref{eq:20})
with $\hat{Q}_{2}$ and
$Q_{2}^{(i)}$ given by Eqs.(\ref{eq:38}) and (\ref{eq:39}):
\be
-i\left[\tilde{\Lambda}(1)+\omega_{0}(1)\right]G_{\hat{M}m}^{(0)}(12)
=\delta (12)~~~.
\ee
and
\be
i\left[\Lambda (1)-\omega_{0}(1)\right]G_{2}^{(0)}(12)
=-\int
d\bar{1} ~\Pi_{0} (1\bar{1})G_{\hat{M}m}^{(0)}(\bar{1}2)
{}~~~.
\label{eq:41}
\ee
The equation for the response function has the solution
\be
G_{\hat{M}m}^{(0)}(r,t_{1}t_{2})=-i\theta (t_{2}-t_{1})R(t_{2}t_{1})
\frac{e^{-\frac{r^{2}}{4(t_{2}-t_{1})}}}{[4\pi (t_{2}-t_{1})]^{d/2}}
=G_{m\hat{M}}^{(0)}(r,t_{2}t_{1})
\ee
where we have introduced
\be
R(t_{1}t_{2})=e^{\int_{t_{2}}^{t_{1}}~d\tau\omega_{0} (\tau )}  ~~~.
\label{eq:43}
\ee
The equation for the correlation function,
Eq.(\ref{eq:41}), can be integrated up into the symmetric form,
suppressing  the step functions in time,
\be
G_{2}^{(0)}(r,t_{1}t_{2})=R(t_{1}t_{0})R(t_{2}t_{0})
\int ~\frac{d^{d}q}{(2\pi )^{d}}e^{i\vec{q}\cdot\vec{r}}\tilde{g}(q)
e^{-2q^{2}T}
\ee
where
$T=\frac{t_{1}+t_{2}}{2}$ and $\tilde{g} (q)$ is the Fourier transform of
$g(r)$.
We focus on the long-time limit where in the integral over wavenumber we can
replace $\tilde{g}(q)\rightarrow \tilde{g}(0)$ to leading order in
powers of $1/T$
and obtain for large $T$:
\be
G_{2}^{(0)}(r,t_{1}t_{2})=R(t_{1}t_{0})R(t_{2}t_{0})\tilde{g}(0)
\frac{e^{-r^{2}/8T}}{(8\pi T)^{d/2}} ~~~.
\ee
If
this system is to order, $<m^{2}(1)>$ is to grow large, then we
must choose, for large time $t$,
$\omega_{0} (t)=2\xi (t)\phi_{0}(t)=\frac{\omega}{t}$
where $\omega$ is a constant\cite{cutoff} we will determine.
Then,
$R(t_{1}t_{2})$ defined by Eq.(\ref{eq:43}), is given by
$R(t_{1}t_{2})=\left(\frac{t_{1}}{t_{2}}\right)^{\omega}$
and
\be
G_{2}^{(0)}(r,t_{1}t_{2})=\tilde{g}(0)
\left(\frac{t_{1}}{t_{0}}\right)^{\omega}
\left(\frac{t_{2}}{t_{0}}\right)^{\omega}
\frac{e^{-r^{2}/8T}}{(8\pi T)^{d/2}}~~~.
\ee
If we are to have a self-consistent scaling equation then the
autocorrelation function $(r=0)$, at large equal times $t_{1}=t_{2}=t$ must
satisfy
\be
G_{2}^{(0)}(0,tt)=S^{(0)}(t)=A_{0}t
=t^{2\omega-d/2}\frac{1}{(t_{0})^{2\omega}}
\frac{\tilde{g}(0)}{(8\pi )^{d/2}}
\ee
which fixes the exponent
$\omega=\frac{1}{2}(1+\frac{d}{2})$
and the amplitude $A_{0}$.
This determination of $\omega$
must be repeated order by order in perturbation
theory.
Returning to the correlation function, we can eliminate
$\tilde{g}(0)$ using the
expression in terms of the amplitude and obtain
$G_{2}^{(0)}(r,t_{1}t_{2})=\sqrt{S(t_{1})S(t_{2})}\Phi_{0}(t_{1}/t_{2})
f_{0}(x)$
where $\Phi_{0}(t_{1}/t_{2})=\left(\frac{\sqrt{t_{1}t_{2}}}{T}
\right)^{d/2}$ and $f_{0}(x)=e^{-x^{2}/2}$.
One can read off from these results the nonequilibrium exponent
$\lambda =\frac{d}{2}$
given by the OJK result.

Going forward,
it is easy to find that the next nonzero
contributions to $Q_{2}$ and $\hat{Q}_{2}$ is of ${\cal O}(2)$ and given by
\be
Q_{2}^{(i)}(12)^{(2)}=-\nabla_{i}^{(1)}\left[
\frac{\omega_{1}(1)}{3!}G_{4}(1112)\right]
\ee
\be
\hat{Q}_{2}(12)^{(2)}=-
\frac{\omega_{1}(1)}{2}G_{\hat{M}mmm}(1112) ~~~.
\ee
These require that we
evaluate the four point cumulants at the leading first
order and obtain
\be
G_{\hat{M}mmm}(1234)=G_{\hat{M}m}^{(0)}(1\bar{1})[-i\omega_{1}(\bar{1})]
\ee
\be
\times\left[G_{\hat{M}m}^{(0)}(\bar{1}2)G_{2}^{(0)}(\bar{1}3)G_{2}^{(0)}(\bar{1}4)
+G_{\hat{M}m}^{(0)}(\bar{1}3)G_{2}^{(0)}(\bar{1}2)G_{2}^{(0)}(\bar{1}4)+G_{\hat{M}m}^{(0)}(\bar{1}4)G_{2}^{(0)}(\bar{1}2)G_{2}^{(0)}(\bar{1}3)\right]
\nonumber
\ee
\be
G_{4}(1234)=G_{2}^{(0)}(1\bar{1})[-i\omega_{1}(\bar{1})]\Bigg[
G_{\hat{M}m}^{(0)}(\bar{1}2)G_{2}^{(0)}(\bar{1}3)G_{2}^{(0)}(\bar{1}4)
\ee
\be
+G_{\hat{M}m}^{(0)}(\bar{1}3)G_{2}^{(0)}(\bar{1}2)G_{2}^{(0)}(\bar{1}4)
+G_{\hat{M}m}^{(0)}(\bar{1}4)G_{2}^{(0)}(\bar{1}2)G_{2}^{(0)}(\bar{1}3)\Bigg]
\nonumber
\ee
\be
+G_{m\hat{M}}^{(0)}(1\bar{1})
(-i\omega_{1}(\bar{1}))G_{2}^{(0)}(\bar{1}2)G_{2}^{(0)}(\bar{1}3)G_{2}^{(0)}(\bar{1}4)
\nonumber
\ee
where we integrate over the repeated barred index.
Notice that $G_{4}$ is properly symmetric
under interchange of its labels.  Inserting these results back into the
equations for the two point quantities we have for the response
function
\be
G_{\hat{M}m}(12)=G_{\hat{M}m}^{(0)}(12)
+\int d\bar{1}d\bar{2}G_{\hat{M}m}^{(0)}(1\bar{1})\Sigma_{\hat{M}m}(\bar{1}\bar{2})
G_{\hat{M}m}^{(0)}(\bar{2}2)
\ee
amd the lowest order self-energy contribution is given by
\be
\Sigma_{\hat{M}m}(12)=\frac{1}{2}[-i\omega_{1}(1)]
G_{\hat{M}m}^{(0)}(12)G_{2}^{(0)}(12)^{2}[-i\omega_{1}(2)]~~~.
\ee
The correlation function is then given to second order by
\be
G_{2}(12)=G_{2}^{(0)}(12)+G_{2}^{(2,1)}(12)+G_{2}^{(2,1)}(21)
+G_{2}^{(2,2)}(12)
\ee
where
\be
G_{2}^{(2,1)}(12)=\int d\bar{1}d\bar{2} G_{2}^{(0)}(1\bar{1})
\Sigma_{\hat{M}m}(\bar{1}\bar{2})G_{\hat{M}m}^{(0)}(\bar{2}2)
\ee
and
\be
G_{2}^{(2,2)}(12)=
-\int d\bar{1}d\bar{2}G_{m\hat{M}}^{(0)}(1\bar{1})\Pi^{(2)}(\bar{1}\bar{2})
G_{\hat{M}m}^{(0)}(\bar{2}2)
\ee
where the self-energy is the same as for the response function and
\be
\Pi^{(2)}(12)=-\frac{1}{3!}[-i\omega_{1}(1)]G_{2}^{(0)}(12)^{3}
[-i\omega_{1}(2)]~~~.
\ee
The spatial integrals giving
$G_{2}^{(2,1)}$ and $G_{2}^{(2,2)}$
can be evaluated in $d$-dimensions
since they involve
products of displaced gaussians.  After rescaling the internal
time integrations $\bar{t}_{1}=Ty_{1}$ and $\bar{t}_{2}=Ty_{2}$ we
obtain
\be
G_{2}^{(2,1)}(12)=\sqrt{S(1)S(2)}2^{d-1}
\omega^{2}\Phi_{0}(t_{1}/t_{2})
J_{1}(x,t_{1}/T,T)
\ee
\be
J_{1}(x,t_{1}/T,T)=
\int_{t_{0}/T}^{t_{1}/T}dy_{1}\int_{t_{0}/T}^{y_{1}}dy_{2}
\frac{y_{1}^{d/2-1}y_{2}^{d/2-1}}
{[(y_{1}+y_{2})(3y_{1}-y_{2}-(y_{1}-y_{2})^{2})]^{d/2}}
e^{-\frac{1}{2}\frac{3y_{1}-y_{2}}{3y_{1}-y_{2}-(y_{1}-y_{2})^{2}}x^{2}}
\nonumber
\ee
\be
G_{2}^{(2,2)}(12)=\sqrt{S(1)S(2)}\frac{2^{d-1}}{3}
\omega^{2}\Phi_{0}(t_{1}/t_{2})
J_{2}(x,t_{1},t_{2})
\ee
\be
J_{2}(x,t_{1},t_{2})=
\int_{t_{0}/T}^{t_{1}/T}dy_{1}\int_{t_{0}/T}^{t_{2}/T}dy_{2}
\frac{y_{1}^{d/2-1}y_{2}^{d/2-1}}
{[(y_{1}+y_{2})^{2}(3-y_{1}-y_{2})]^{d/2}}
e^{-\frac{1}{2}\frac{3}{3-y_{1}-y_{2}}x^{2}} ~~~.
\nonumber
\ee
The first thing we should do with these results is look at the contribution
at this order to the on site equal-time $t_{1}=t_{2}=t$
correlation function.
The integrals $J_{1}(0,1,t)$ and $J_{2}(0,t,t)$ are
logarithmically divergent as $t\rightarrow\infty$
and we have
\be
S(1)=S^{(0)}(t)\left[1+\omega^{2}2^{d}\left(K_{d}+
\frac{M_{d}}{3^{d/2+1}}\right) ln(t/t_{0})+\cdots \right]
\label{eq:69}
\ee
where $K_{d}$ is defined by Eq.(\ref{eq:7}) and $M_{d}$ is
defined by Eq.(\ref{eq:8}).
Remembering that at lowest order
$S^{(0)}(t)=A_{0}t^{2\omega-d/2}$
and after exponentiation of the $ln(t/t_{0})$ term in
Eq.(\ref{eq:69}) we find that to obtain
$S(t)=At$, we require that $\omega$ be given by
Eq.(\ref{eq:6}).

The $t_{0}\rightarrow 0$ singularities in $J_{1}$ and $J_{2}$
can be regulated by turning our
attention from $G_{2}(12)$ to the scaled quantities $\Phi (t_{1}/t_{2})$
and $f(x,t_{1}/t_{2})$ which appear in Eq.(\ref{eq:3}):
\be
f(12)=\frac{G_{2}(12)}{\sqrt{S(1)S(2)}}
=\Phi (t_{1}/t_{2})f(x,t_{1}/t_{2}) ~~~.
\label{eq:70}
\ee
Looking at the second order contributions to $\Phi (t_{1}/t_{2})$ we
find for $t_{1}\gg t_{2}$ or $t_{2}\gg t_{1}$ there are contributions
which are logarithmically divergent  in the quantity
$\sqrt{t_{1}t_{2}}/T$
which can be exponentiated to obtain the result given by
Eq.(\ref{eq:4}).
$f(x,t_{1}/t_{2})$
is an analytic function of $x$.  One can  easily work out, for example,
the lowest order
term in the power
series expansion in $x^{2}$  for $f(x,t_{1}/t_{2})$ which is regular for all
$d$.  It is this term which gives the coefficient in Porod's law\cite{porod}.
There is one last regularization that must be carried out before the
perturbation theory expression for the correlation function can be
used for all values of $x$. If we look at large $x$ we find that
$f(x,t_{1}/t_{2})$ is logarithmically divergent in $1+x^{2}$
which, on exponentiation leads to the expression given by
Eq.(\ref{eq:9}).

We turn next to the connection between the correlation function for the
auxiliary field $m$ and the order parameter correlation function.
As discussed earlier,
since $u(1)$ vanishes exponentially for large
$|m(1)|$, then the averages over these fields are down by a factor of
$L^{-2}$ relative to the averages over the field $\sigma$ and in the
scaling regime
$C(12)=\left<\sigma (1)\sigma (2)\right>$.
The perturbation expansion for this quantity follows closely the
development for the one-point quantity except we must treat the two-
point generalization of $P_{h}(x_{1},1)$.  One can then show
\be
C(12)=\frac{2}{\pi}~sin^{-1}~f(12)
+{\cal O}(3)
\ee
where $f(12)$ is defined by Eq.(\ref{eq:70}).
Since we have worked out $f(12)$ to second order previously, we
have consistently to this order, the result given by Eq.(\ref{eq:3}).

Clearly one can go to higher order in this perturbation theory and
introduce more sophisticated graphical methods.  It is interesting
to ponder  the implications of the
ideas developed here for the conserved and
vector order parameter cases.

\centerline{Acknowledgements}

I thank Dr. R. Wickham for useful comments.
This work was supported in part by the MRSEC Program of the National Science
Foundation under Award Number DMR-9400379.
%
%


\begin{thebibliography}{99}
\bibitem{quan} The quantities $\Delta\Phi (t_{1}/t_{2})$ and
$\Delta f (x,t_{1}/t_{2})$ have been determined but the expressions for
them are long and not very illuminating.
\bibitem{second}  The nature of the expansion is discussed below.
Convergence properties, if any, remain to be investigated.
\bibitem{conservation} The quantity
$\rho_{i} $ is the scalar order parameter defect density.  Since we
expect that $m$ and $\psi$ share the same zeros, we have
$\rho_{i}=\delta (\psi )\nabla_{i} \psi$.  As discussed in
G. F. Mazenko, Phys. Rev. Lett. {\bf 78}, 401 (1997), $\rho_{i} $
is conserved for $n=d=1$.  It is easy to see that this holds for all
$d$ for the scalar case.
\bibitem{TUG2}This
is just the transformation introduced in Ref.(\onlinecite{TUG})
and used in Ref.(\onlinecite{BH}).
\bibitem{causality}
One point to keep in
mind is the well known property of the MSR method that the
equal-time correlation functions with hatted fields must vanish if one is to
statisfy causality:
$G_{\hat{M}m}(11)=0$.
\bibitem{cutoff}  One can regulate the theory at short times by introducing the
time cutoff $t_{c}$ via $\omega_{0}(t)=
\omega /(t+t_{c})$.
\end{thebibliography}
\end{document}